\documentclass[12pt,english]{article}
\usepackage[T1]{fontenc}
\usepackage[latin1]{inputenc}

\makeatletter

\providecommand{\LyX}{L\kern-.1667em\lower.25em\hbox{Y}\kern-.125emX\@}

\usepackage{amsfonts}
\usepackage{epsfig}
\usepackage{psfig}
\usepackage{graphicx}
\usepackage{amsmath}
\usepackage{babel}
\makeatother
\begin{document}

\title{A note on the Lorentz force, magnetic charges and the Casimir effect }

\author{C. Farina%
\footnote{e-mail: farina@if.ufrj.br%
}, F. C. Santos%
\footnote{e-mail: filadelf@if.ufrj.br%
} and A. C. Tort%
\footnote{e-mail:tort@if.ufrj.br; present address: Institut d'Estudis Espacials
de Catalunya (IEEC/CSIC) Edifici Nexus 201, Gran Capit\`{a} 2-4,
08034 Barcelona, Spain; e-mail address: visit11@ieec.fcr.es%
}\\
 Instituto de F\'{\i}sica\break \\
 Universidade Federal do Rio de Janeiro\break \\
 Cidade Universit\'{a}ria - Ilha do Fund\~{a}o - Caixa Postal 68528\\
 21941-972 Rio de Janeiro RJ, Brasil.}

\date{\today}

\maketitle
\begin{abstract}
We show that in order to account for the repulsive Casimir effect
in the parallel plate geometry in terms of the quantum version of
the Lorentz force, virtual surface densities of magnetic charge and
currents must be introduced. The quantum version of the Lorentz force
expressed in terms of the correlators of the electric and the magnetic
fields for planar geometries yields then correctly the Casimir pressure. 
\end{abstract}
\noindent PACS: 11. 10. -z; 12. 20. -m

\section{Introduction}

Since its prediction by Casimir in 1948, the Casimir effect \cite{Casimir1948}
has been the object of an increasing theoretical and experimental
investigation. This is due to its growing recognition as a fundamental
feature of quantum field theory and also to its importance in elementary
particle physics, cosmology and condensed matter physics as well as
its practical and decisive role in nanotechnology. For an introduction
to this remarkable effect, see for example \cite{Milonni1994}; for
an updated review of the recent research and applications see \cite{MosteTrunov}
and references therein. 

Even in simple examples the Casimir interaction can exhibit surprising
features. Some time ago Gonzales \cite{Gonzales}, among other things,
correctly pointed out that an alternative computation of the Casimir
force between two perfectly conducting plates can be carried out starting
from the consideration of the Lorentz force acting on the plates.
The reason is that in the context this is the only force that could
act on a metallic plate and therefore one should be able to obtain
Casimir's result from this physical fact. Here we will develop further
this point of view and consider its consequences when applied to Boyer's
variant of the standard Casimir effect in which one of the conducting
plates is replaced by a magnetically permeable one \cite{Boyer74}.
We will show that when applied to this particular case, Gonzales'
conception of the Casimir interaction leads to the introduction of
virtual magnetic charges and currents.

\section{The standard Casimir effect}

In order to state clearly our point of view we begin by considering
the standard experimental setup proposed by Casimir which consists
of two infinite perfectly conducting parallel plates kept at a fixed
distance $a$ from each other. We will choose the coordinates axis
in such a way that the $OZ$ direction is perpendicular to the plates.
One of the plates will be placed at $z=0$ and the other one at $z=a$.
Classicaly, the Lorentz force per unit area on, say, the conducting
plate at $z=a$ is given by \begin{equation}
\vec{f}_{e}=\frac{1}{2}\sigma _{e}\vec{E}+\frac{1}{2c}\vec{K}_{e}\times \vec{B},\label{LorentzElec}\end{equation}
where $\sigma _{e}$ is the electric charge density and $\vec{K}_{e}$
is the electric current surface density. The boundary conditions on
the electric and the magnetic fields on the plate are: the tangential
components $E_{x}$ and $E_{y}$ of the electric field and the normal
component $B_{z}$ of the magnetic field must be zero on the plate.
Under the boundary conditions imposed on the field at $z=a$, however,
it is easily seen that the resultant classical Lorentz force is perpendicular
to the conducting plate. We expect the quantum version of Eq. (\ref{LorentzElec})
to show the same feature and ultimately to be the source of the Casimir
pressure between the two conducting plates. The electric charge and
current densities on the conducting plate are related to the fields
through\begin{equation}
\hat{n}\cdot \vec{E}=4\pi \sigma _{e},\label{Maxwell 1}\end{equation}
\begin{equation}
\hat{n}\times \vec{B}=\frac{4\pi }{c}\vec{K}_{e},\label{Maxwell 2}\end{equation}
where $\hat{n}$ is the normal to the plate under consideration. The
quantum version of Eq. (\ref{LorentzElec}) reads\begin{equation}
\left\langle \vec{f}_{e}\right\rangle _{0}=\frac{1}{8\pi }\left\langle \vec{E}^{2}-\vec{B}^{2}\right\rangle _{0}\, \hat{n},\end{equation}
and can be obtained by combining the vacuum expectation value of Eqs.
(\ref{LorentzElec}), (\ref{Maxwell 1}), and (\ref{Maxwell 2}).
In order to proceed -- from now on we depart from Ref. \cite{Gonzales}
-- we need to evaluate the vacuum expectation value of the quantum
operators $E_{i}\left(\vec{r},t\right)E_{j}\left(\vec{r},t\right)$,
$B_{i}\left(\vec{r},t\right)B_{j}\left(\vec{r},t\right),$ and $E_{i}\left(\vec{r},t\right)B_{j}\left(\vec{r},t\right)$.
The evaluation of these correlators depends on the specific choice
of the boundary conditions. A regularisation recipe is also necessary,
for these objects are mathematically ill-defined. Regularisation recipes
vary from the relatively simple cutoff method employed by Casimir
himself \cite{Casimir1948} to the sophisticated and mathematical
elegant generalised zeta function techniques, see Refs. \cite{Reuter&Dittrich1985}
for an introdution to these techniques. Here we will make use of the
results and send the reader to the relevant references. The electric
field correlators for a pair of perfectly conducting plates are given
by \cite{LüRavndal85,Bordag85,Barton90,Jandiretal00}\begin{equation}
\langle E_{i}(\vec{r},t)E_{j}(\vec{r},t)\rangle _{0}=\left(\frac{\pi }{a}\right)^{4}{\frac{2}{3\pi }}\left[\frac{\left(-\delta ^{\Vert }+\delta ^{\bot }\right)_{ij}}{120}+\delta _{ij}F(\xi )\right]\label{ECorrCasimir2}\end{equation}
The function $F\left(\xi \right)$ with $\xi :=\pi z/a$ is defined
by \begin{equation}
F\left(\xi \right):=-\frac{1}{8}\frac{d^{3}\, }{d\xi ^{3}}\frac{1}{2}\cot \left(\xi \right),\end{equation}
 and its expansion about $\xi =\xi _{0}$ is given by \begin{equation}
F\left(\xi \right)\approx \frac{3}{8}\left(\xi -\xi _{0}\right)^{-4}+\frac{1}{120}+O\left[\left(\xi -\xi _{0}\right)\right]^{2}.\label{Fapp}\end{equation}
 Notice that due to the behavior of $F\left(\xi \right)$ near $\xi _{0}=0,\pi $,
strong divergences control the behavior of the correlators near the
plates. The corresponding magnetic field correlators are \begin{equation}
\langle B_{i}(\vec{r},t)B_{j}(\vec{r},t)\rangle _{0}=\left({\frac{\pi }{a}}\right)^{4}{\frac{2}{3\pi }}\left[\frac{\left(-\delta ^{\Vert }+\delta ^{\bot }\right)_{ij}}{120}-\delta _{ij}F(\xi )\right]\, .\label{BCORRCASIMIR}\end{equation}
 A direct evaluation also shows that the correlators $<E_{i}(\vec{r},t)B_{j}(\vec{r},t)\rangle _{0}$
are zero. For calculational purposes it is convenient to consider
a third conducting plate placed perpendicularly to the $OZ$ axis
at $z=\ell $. Consider the plate at $z=a$. The Lorentz force per
unit area on its left side ($\hat{n}=-\hat{z}$) reads\begin{equation}
\left\langle f_{z}^{L}\right\rangle _{0}=-\frac{1}{8\pi }\left\langle \vec{E}^{2}-\vec{B}^{2}\right\rangle _{0}\approx -\frac{3}{16\pi ^{2}\left(z-a\right)^{4}}-\frac{\pi ^{2}}{240\, a^{4}},\end{equation}
where we have also made use of Eq. (\ref{Fapp}). On the other hand,
after simple modifications in Eqs. (\ref{ECorrCasimir2}), (\ref{BCORRCASIMIR})
and (\ref{Fapp}) the Lorentz force on the right side of the plate
($\hat{n}=\hat{z})$ reads \begin{equation}
\left\langle f_{z}^{R}\right\rangle _{0}=\frac{1}{8\pi }\left\langle \vec{E}^{2}-\vec{B}^{2}\right\rangle _{0}\approx \frac{3}{16\pi ^{2}\left(z-a\right)^{4}}+\frac{\pi ^{2}}{240\, \left(\ell -a\right)^{4}}.\end{equation}
Adding the forces on both sides of the plate and setting $\ell \rightarrow \infty $
we obtain the well known result\begin{equation}
\left\langle f_{z}\right\rangle _{0}=\left\langle f_{z}^{R}\right\rangle _{0}+\left\langle f_{z}^{R}\right\rangle _{0}=-\frac{\pi ^{2}}{240\, a^{4}}.\end{equation}
The minus sign means that the probe plate at $z=a$ is attracted towards
the other one at the origin.

\section{The repulsive version of the standard Casimir effect}

Let us now consider an alternative setup to the standard one in which
a perfectly conducting plate is placed at $z=0$ and perfectly permeable
one is placed at $z=a$. This setup was analysed for the first time
by Boyer in the context of stochastic electrodynamics \cite{Boyer74},
a kind of classical electrodynamics that includes the zero-point electromagnetic
radiation, and leads to the simplest example of a repulsive Casimir
interaction. For alternative evaluations see Refs. \cite{Santos&Tort99,Santana2001}.
The boundary conditions now are: \emph{(a)} the tangential components
$E_{x}$ and $E_{y}$ of the electric field as well as the normal
component $B_{z}$ of the magnetic field must vanish on the surface
of the plate at $z=0$; \emph{(b)} the tangential components of $B_{x}$
e $B_{y}$ of the magnetic field as well as normal conponent $E_{z}$
of the electric field must vanish on the surface of the plate at $z=a$.
From the classical point of view the perfectly permeably plate at
$z=a$ poses a problem when we apply Eq. (\ref{LorentzElec}) to it.
This is so because due to the boundary conditions this time the Lorentz
force on either side of the plate is \emph{parallel} to the permeable
plate and therefore the resultant Lorentz force will be also parallel
to the plate. This is a puzzling feature if we wish to describe the
Casimir interaction beteween the plates through the Lorentz force.
Things can be mended, however, if we allow for a virtual surface magnetic
charge density $\sigma _{m}$ and a virtual magnetic charge current
surface density $\vec{K}_{m}$. In this case the modified Lorentz
force per unit area on the permeable plate reads\begin{equation}
\vec{f}_{m}=\frac{1}{2}\sigma _{m}\vec{B}-\frac{1}{2c}\vec{K}_{m}\times \vec{E}.\label{LorentzMag}\end{equation}
The charge and current surface densities are related to the fields
through\begin{equation}
\hat{n}\cdot \vec{B}=4\pi \sigma _{m},\end{equation}
\begin{equation}
\hat{n}\times \vec{E}=-\frac{4\pi }{c}\vec{K}_{m}.\end{equation}
It is easily seen that the modified Lorentz force given by Eq. (\ref{LorentzMag})
combined with the boundary conditions on the permeable plate yields
on either side of the plate a force perpendicular to the plate as
it must be. Proceeding as above we now have\begin{equation}
\left\langle \vec{f}_{m}\right\rangle _{0}=\frac{1}{8\pi }\left\langle \vec{B}^{2}-\vec{E}^{2}\right\rangle _{0}\, \hat{n}\end{equation}
For Boyer's setup the relevant correlators were evaluated in Refs.
\cite{JPA99,Jandiretal00}. The results are\begin{equation}
\left\langle E_{i}\left(\vec{r},t\right)E_{j}\left(\vec{r},t\right)\right\rangle _{0}=\left(\frac{\pi }{a}\right)^{4}\frac{2}{3\pi }\left[\left(-\frac{7}{8}\right)\frac{\left(-\delta ^{\Vert }+\delta ^{\perp }\right)_{ij}}{120}+\delta _{ij}\, G\left(\xi \right)\right]\, ,\label{EcorrBoyer}\end{equation}
 \begin{equation}
\left\langle B_{i}\left(\vec{r},t\right)B_{j}\left(\vec{r},t\right)\right\rangle _{0}=\left(\frac{\pi }{a}\right)^{4}\frac{2}{3\pi }\left[\left(-\frac{7}{8}\right)\frac{\left(-\delta ^{\Vert }+\delta ^{\perp }\right)_{ij}}{120}-\delta _{ij}\, G\left(\xi \right)\right]\, ,\label{BcorrBoyer}\end{equation}
 where\begin{equation}
G\left(\xi \right)=-\frac{1}{8}\frac{d^{3}\, }{d\xi ^{3}}\frac{1}{2\sin \left(\xi \right)}.\end{equation}
 \[
\]
Near $\xi =0$ the function $G\left(\xi \right)$ behaves as \begin{equation}
G\left(\xi \right)=\frac{3}{8}\xi ^{-4}-\frac{7}{8}\, \frac{1}{120}+O\left(\xi ^{2}\right),\label{Gapp1}\end{equation}
 but near $\xi =\pi $ its behavior is slightly different \begin{equation}
G\left(\xi \right)=-\frac{3}{8}\left(\xi -\pi \right)^{-4}+\frac{7}{8}\, \frac{1}{120}+O\left[\left(\xi -\pi \right)^{2}\right].\label{Gapp2}\end{equation}
 Again, a direct calculation shows that $\left\langle E_{i}\left(\vec{r},t\right)B_{j}\left(\vec{r},t\right)\right\rangle _{0}=0$
for this case also. 

To obtain the Casimir force it is convenient to replace the third
plate at $z=\ell $ for a permeable one. It is not hard to see that
if we do so we can use Eqs. (\ref{ECorrCasimir2}) and (\ref{BCORRCASIMIR})
in the region between the two permeables plates with small modifications.
The force on the left side ($\hat{n}=-\hat{z})$ of the permeable
plate at $z=a$ then reads\begin{equation}
\left\langle f_{m,z}^{L}\right\rangle _{0}=-\frac{3}{16\pi ^{2}\left(z-a\right)^{4}}+\frac{7}{8}\times \frac{\pi ^{2}}{240\, a^{4}},\end{equation}
 and the force on the right side is\begin{equation}
\left\langle f_{m,z}^{R}\right\rangle _{0}=\approx \frac{3}{16\pi ^{2}\left(z-a\right)^{4}}+\frac{\pi ^{2}}{240\, \left(\ell -a\right)^{4}}.\end{equation}
As before we add the forces on each side and set $\ell \rightarrow \infty $
to obtain the repulsive Casimir force per unit area\begin{equation}
\left\langle f_{m,z}\right\rangle _{0}=\left\langle f_{m,z}^{L}\right\rangle _{0}+\left\langle f_{m,z}^{R}\right\rangle _{0}=\frac{7}{8}\times \frac{\pi ^{2}}{240\, a^{4}},\label{Repulsiveforce}\end{equation}
in agreement with Boyer \cite{Boyer74}. Notice that this time the
force per unit area is repulsive. Notice also that in both cases the
divergent pieces cancel out; these cancellations yield finite Casimir
energies \cite{LüRavndal85,Barton90,JPA99,Jandiretal00}.

\section{Conclusions}

As we can see the introduction of virtual magnetic charges and currents
can account for Boyer's variant of the Casimir effect in terms of
the quantum version of the Lorentz force. Of course, for other geometries
and boundary conditions, for example a perfectly conducting cube,
the Casimir force can be repulsive and accountable for by the usual
virtual electric charges and currents. On the other hand, it is plausible
to state that whenever we have an ideal magnetically permeable wall
as one of the confining surfaces a qualitative analysis of the interaction
between the zero point electromagnetic fields and this confining surface,
which is modelled by appropriate boundary conditions, will show the
need of introducing virtual magnetic charges and currents. The introduction
of these charges and currents avoids the need of constructing a model
of the Casimir interaction based on an appropriate distribution of
amperian currents, a considerably harder task. 

As a final remark we observe that in the framework of the usual cavity
QED, the result given by Eq. (\ref{Repulsiveforce}) is obtained by
first evaluating the confined vacuum renormalized energy and followed
by a variation of the volume of the confining region. It can be easily
shown, however, that in terms of partition functions and free energies,
Boyer's setup is mathematically equivalent to the difference between
two standard setups, one with the distance between the plates equal
to $2a$ and the other one with the distance between the plates equal
to $a$, see Ref. \cite{Santos&Tort99}. This can be also easily proved
for the Casimir pressure at zero and finite temperature.

\section*{Acknowledgments}

The authors are greatful to P A M Neto and M V Cougo-Pinto for valuable
observations. One of the authors (A C T) wishes to acknowledge E Elizalde
and the hospitality of the Institut d'Estudis Espacials de Catalunya
(IEEC/CSIC) and Universitat de Barcelona, Departament d'Estrutura
i Constituents de la Mat\`{e}ria where this work was completed, and
also the financial support of Consejo Superior de Investigaci\'on
Cient\'{\i}fica (Spain) and CAPES, the Brazilian agency for faculty
improvement, Grant BEX 0682/01-2.

\end{document}